\documentclass[10pt,a4paper,authoryear]{article} 
\usepackage[utf8]{inputenc}
\usepackage[english]{babel}
\usepackage{amsmath,amssymb}
\usepackage{graphicx}
\usepackage{geometry}
\usepackage{authblk}
\usepackage{breakcites}

\usepackage{oplotsymbl} 
\usepackage{fdsymbol} 

\title{A biophysical approach to the design of networks of communication systems}
\author{Rodrigo Almeida,   Ana Filipa Valente,  Rui Dil\~ao}

\affil{University of Lisbon, Instituto Superior T\'ecnico, Dep. of Physics\\
Av. Rovisco Pais, 1049-001 Lisbon, Portugal}

\date{}

\begin{document}

\maketitle

\begin{abstract}
Inspired by the growth dynamics of the protist \textit{Physarum polycephalum}, we employ a formalism that describes adaptive, incompressible Hagen-Poiseuille flows on channel networks to identify graphs connecting different nodes within Euclidean space. These graphs are either suboptimal or optimal with respect to their length. Occasionally, we derive graph tree configurations that are topologically equivalent to Steiner trees. This methodology can be utilised to assist in making decisions regarding the design of communication networks, such as fibre webs, motorways, or railway networks. As a demonstration of the practicality of this approach, we explicitly apply this framework to the Portuguese railway network.
\end{abstract}



\section{Introduction}

Multiple classes of equations employing Hagen-Poiseuille (H-P) flows have been developed to elucidate the growth of the protist \textit{Physarum polycephalum}, commonly referred to as slime mould. Despite the absence of a neural system or centralised control, this organism demonstrates sophisticated behaviour, such as the ability to select the shortest pathway between food sources within mazes \cite{maze}. When presented with multiple food sources, \textit{Physarum} can develop networks exhibiting comparable efficiency, fault tolerance, and cost-effectiveness to artificial networks \cite{tero3, tero_tokyo}. Various biological experiments suggest that the network tree structures of slime mould often resemble Steiner trees. Algorithms based on adaptive H-P type flows have been devised to interpret both \textit{Physarum} network formation and Steiner tree development \cite{tero_steiner, sun_physarum_steiner, rodrigo_tese, hsu_steiner_physarum}. The Steiner tree problem pertains to network optimisation within graphs and holds substantial relevance in the fields of physics, biology, and engineering \cite{sun_physarum_drugs, travelling_salesman}. 

The survival of multicellular and multinucleated organisms depends on their vascular network. A vein network must effectively transport and distribute resources and nutrients throughout the body to ensure optimal survival, providing short pathways between critical locations. To enhance transport efficiency, a biological network should be as concise as possible to minimise the costs associated with network construction and energy dissipation. Additionally, biological networks must be robust and possess redundancy to prevent failure in the event of damage to a channel. In the case of \textit{Physarum}, the adaptability of its vein network is essential for foraging and survival.

Previous models, based on the phenomenology of adaptive H-P flows, have investigated the formation and optimisation of network pathways in \textit{Physarum} by utilising a deterministic formalism to solve systems of ordinary differential equations on graphs \cite{transport1, rodrigo_artigo}. These graphs cover bounded regions within Euclidean space. The vertices of the graph can denote branching points of veins or edges, as well as sources or sinks of an incompressible fluid. Given the initial inflows and outflows at the sources and sinks, along with the total fluid volume within the network, and under the assumption that the radius of the network veins is elastic and capable of adaptation as a function of flux, the flow on the graph achieves a steady state whereby certain veins transport fluid while others do not. By maintaining constant inflows and outflows over time, a steady state is attained with a fixed subgraph or tree geometry. The steady states are inherently non-unique and depend on the initial conditions of the adapted conductivities. A notable characteristic of this approach is that if one of the connections in the steady-state subgraph is deactivated, an alternative link within the graph is formed, exemplifying the adaptability and resilience inherent in this process ef {Val2}. This geometric construct emulates several natural systems that are sensitive to the spatial distribution of sources, sinks, and local flows.

This paper explores extremal length graphs connecting $n$ points in Euclidean space by utilising the properties of adaptive incompressible H-P flows that conserve fluid volume within the network \cite{rodrigo_tese,rodrigo_artigo, Val, Val2}. To simulate communication systems and incorporate a level of variability capable of representing the resilient characteristics of natural flows, we have introduced multiple stochastic actualisation processes over time, resulting in suboptimal or optimal solutions for graph networks. In all tested cases, consistent attainment of steady solutions- topologically analogous to Steiner tree solutions- has been observed, which can subsequently be further optimised with respect to graph length. Additionally, suboptimal solutions have been derived for an idealised Portuguese railway network that interconnects all major cities, as well as land and sea borders.

\section{Finding paths of extremal lengths}

Consider a point lattice within a bounded region of Euclidean space of dimension $\mathcal{N}$. The point lattice can be represented by an undirected graph $\mathcal{G} = (\mathcal{V},E)$, where $\mathcal{V}$ is the set of vertices and $E$ is the set of edges. Given a bounded domain $\mathcal{D}$ of Euclidean space, we can construct the graph  $\mathcal{G}$ through a Delaunay triangulation of a set of points with coordinates $(x_i^1,...,x_i^\mathcal{N})$, with $i = 1, ..., N$. Suppose $N$ is sufficiently large and well distributed within $\mathcal{D}$. In that case, we can make the area of triangles sufficiently small and approximate any Euclidean path by a collection of straight edges $(i,j)\in E$ connecting nodes $i$ and $j$. The graph $\mathcal{G}$ is connected, having a single component. The formalism established here is very general, though we assume that $\mathcal{N}=2$. 

From the graph $\mathcal{G}$, we choose $n<<N$ points.  The set of $n$ points may represent the sources and sinks of a fluid injected into the network. Thus, the graph illustrates a network of channels through which an incompressible fluid circulates. Fluid is injected from the sources and exits through the sinks. The inflow and outflow are balanced, resulting in zero net flow.
Our primary objective is to investigate these types of flows, further assuming that the edges of $\mathcal{G}$ can adjust such that they are capable of either conducting or not conducting the fluid.

The question is, under what conditions does this flow adapt to an optimised network comprising a subset of conducting edges that connect the $n$ points of the graph $\mathcal{G}$? 

\begin{figure}[h]
\centering
 \includegraphics[width=0.45\textwidth]{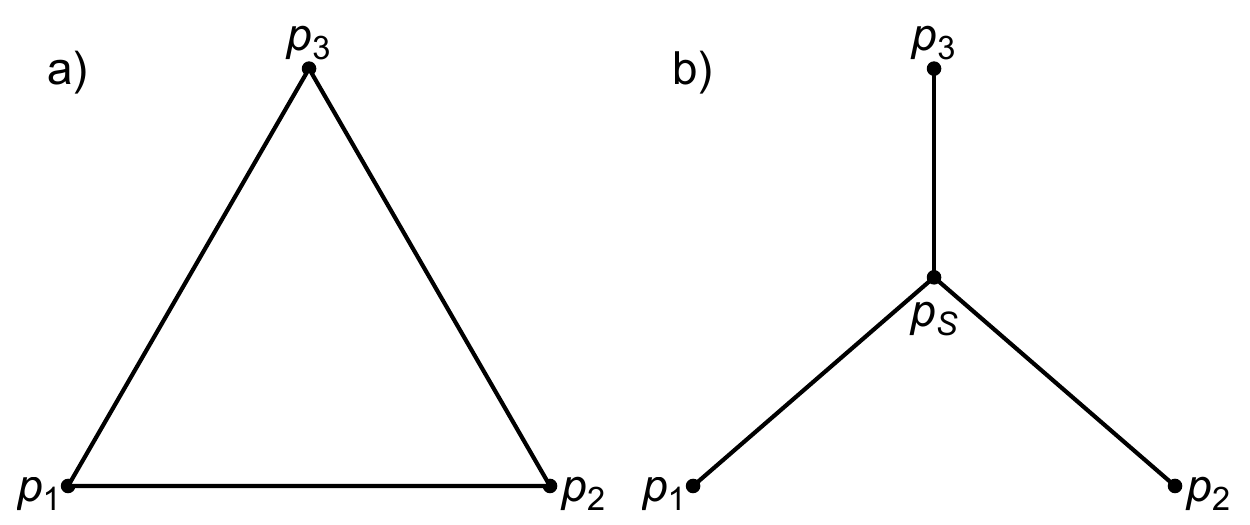}
\caption{Two possible graphs connecting the three points \eqref{points}. a) A graph connecting the three points in the plane. b) The Steiner tree of minimal length connecting the three points in the plane. In this case, the tree has an additional point --- the Steiner point $p_S$.}
\label{fig:triangle_theo}
\end{figure}

For example, consider the case where $n=3$ in a bounded region of the Euclidean plane. The three points are arranged in a triangular configuration. Two possible graphs connect the three points, as illustrated in figure \ref{fig:triangle_theo}. The coordinates of the three points are
\begin{equation}
p_1 = (0,0),\ \ p_2 = (1,0),\ \  p_3 = \left(\frac{1}{2},\frac{\sqrt{3}}{2}\right). 
\label{points}
\end{equation}
A tree or closed graph can connect these three points, as depicted in figure \ref{fig:triangle_theo}.

By definition, the Steiner tree of a set of points on the Euclidean plane is a tree of minimal length that connects all the points. The graph in figure~\ref{fig:triangle_theo}a) has length $L_a = 3$, and in b), $L_b = 3\sqrt{7}/4\approx1.98$. The graph in figure \ref{fig:triangle_theo}b) represents the tree of minimal lengths connecting the three points --- the Steiner tree connecting these points. In the Steiner tree, there exists a fourth point with coordinates $p_S = \left(\frac{1}{2},\frac{\sqrt{3}}{4}\right)$. The challenge in constructing Steiner trees is that, in some instances, there are additional points --- Steiner points --- that, along with the initial points, span the tree of minimal length.

There are two issues in constructing a Steiner tree. The first is determining the number of Steiner points that organise the Steiner tree. The second step is to identify which edges connect the initial data points to the Steiner points. Suppose we know these for a configuration of $n$ points. In that case, the coordinates of the Steiner points can be easily obtained by solving linear systems of equations involving the unknown coordinates of the Steiner points \cite{Val}.

The tree connecting the three points can also represent numerous physical, social, or biological systems. For instance, it can serve as a model of a road network linking three cities. Assume each road is utilised by travellers moving from city $i$ to city $j$, and the number of travellers travelling from $i$ to $j$, where $i\neq j$, remains approximately the same across every $(i,j)$ pair. In the case of the road network with a Steiner tree geometry depicted in the graph of figure 
\ref{fig:triangle_theo}b), the mean distance between two cities is $L_b = 1.98$, whereas for the network shown in figure \ref{fig:triangle_theo}a), the mean distance is $L_a = 1$. From a communication standpoint, the configuration in figure \ref{fig:triangle_theo}a) proves more advantageous for users owing to its reduced transportation costs and enhanced communication efficiency. Nonetheless, its construction cost is higher in comparison to the Steiner tree solution. In section~\ref{sec4}, we establish the optimisation criteria for analysing communication systems.

\section{Adaptive incompressible Hagen-Poiseuille flows on graphs} \label{sec:adaptive-hp}

Certain systems, including blood vessels or \textit{Physarum}, can be mathematically modelled by a class of equations that describe the steady-state flow of viscous, incompressible fluids within networks composed of straight channels featuring multiple sources and sinks, as well as adaptive conductivities \cite{rubinow_book,  transport1, Val2}. The flow resembles Hagen-Poiseuille flow, and the volume of fluid within the network remains constant \cite{rodrigo_artigo}.

A network of veins is represented by an undirected graph subset of $\mathcal{G} = (\mathcal{V},E)$. 
The network has $K$ sources and $R$ sinks situated at fixed nodes, and $n=K+R$. The fluid flux entering or exiting the network at node $j$ is denoted by $ S_j$. If node $j$ hosts a source, then $S_j > 0$; if it hosts a sink, then $S_j < 0$; otherwise, $S_j = 0$. Given the incompressibility of the fluid,
$$
    \sum_{j = 1}^n S_j = \sum_{j:\text{sources}}S_j + \sum_{j:\text{sinks}}S_j = 0,
$$
and the volume of fluid in the system remains constant.

The steady state of the flow within the network is determined by Kirchhoff's law, which stipulates that flux is conserved at each node $j$,
\begin{equation} \label{eq:kirchhoff}
    \sum_{j:(i,j)\in E} Q_{ij} =\sum_{j:(i,j)\in E} D_{ij}\frac{(p_i-p_j)}{L_{ij}} =S_i \quad , i = 1, ..., N,
\end{equation}
where $p_i$ denotes the pressure at node $i$, $L_{ij}$ is the length of the graph edge linking nodes $i$ and $j$, and $D_{ij}$ is the conductivity of the edge $(i,j)$.

The edges represent elastic cylindrical channels capable of expanding or contracting transversely relative to the flow. The edge connecting node $i$ to node $j$ possesses a fixed length $L_{ij}$ and a variable radius $r_{ij}$; the edge $(i,j)$ has (variable) conductivity $D_{ij}$ defined by $D_{ij} = \pi r_{ij}^4/8\eta$, with $\eta$ the dynamic viscosity of the fluid. The volume of the channel connecting node $i$ to node $j$ is given by $V_{ij} = \pi r_{ij}^2 L_{ij} = \sqrt{8\pi\eta}L_{ij}\sqrt{D_{ij}}$, \cite{Lan}. As the fluid is incompressible, the total volume of fluid within the network remains invariant and is given by 
$$
V = \sum_{(i,j)\in E} V_{ij} = \beta \sum_{(i,j)\in E} L_{ij}\sqrt{D_{ij}},
$$
where $\beta = \sqrt{8\pi\eta}$.

The formalism developed in \cite{rodrigo_artigo} asserts that the conductivity of the network's channels follows the adaptation law.
\begin{equation} \label{eq:dij_adaptation}
    \frac{d}{dt}\sqrt{D_{ij}} = \alpha \frac{g(Q_{ij})}{\sum_{(k,m)\in E} L_{km}\, g(Q_{km})} - \sqrt{D_{ij}} \quad,\quad (i,j)\in E,
\end{equation}
where $g(\cdot)$ is an arbitrary function characterising the elasticity of the channels, $\alpha = V/\beta$, and $t$ denotes a dimensionless time. 
The function $g(\cdot)$ is derived by  minimising the dissipated energy per unit time of the H-P flow, denoted by $\mathcal{P}= \sum_{(i,j)\in E}Q_{ij}^2 L_{ij} / D_{ij}$, with respect to $\sqrt{D_{ij}}$, thereby resulting in
\begin{equation} \label{eq:g_minimum}
    g(Q_{ij}) = Q_{ij}^{2/3}.
\end{equation}

To derive the steady-state solution of the adaptive incompressible H-P flow on a graph, we select a set of $n$ sources and sinks, and we initialise all edges of the graph $\mathcal{G}$ with conductivities $D_{ij}(t=0)>0$. These conductivities remain positive for all edges within $\mathcal{G}$.
To describe the temporal evolution of the radii of the channels/edges of $\mathcal{G}$,  we first solve the linear equations detailed in equation \eqref{eq:kirchhoff} with respect to the pressures. Subsequently, the fluxes $Q_{ij}$ are determined by the same equations \eqref{eq:kirchhoff}. The conductivities of the channels are adapted by equations \eqref{eq:dij_adaptation}-\eqref{eq:g_minimum}. The equations \label{eq:dij_adaptation} are integrated using the Euler method with a time step of $\Delta t$. The process is iterated over time.

Assuming that the source and sink fluxes remain constant over time, the adaptation process converges to a steady state of the channel conductivities. A subgraph of $\mathcal{G}$ is forms, comprising conductivities that exceed a specified threshold. This steady state is dependent on the initial conductivities, and the subgraph may be disconnected, \cite{rodrigo_tese}.

To emulate communication systems in which the roles of sources or sinks fluctuate over time, the conductivity values $D_{ij}$ do not attain a stable value and vary with each iteration. Nonetheless, the geometry of the steady-state graph remains constant. 

To simulate all possible itineraries between the nodes of a network and determine a steady-state path of the embedded subgraph in the initial graph $\mathcal{G}$, we utilise four distinct stochastic algorithms to update the conductivities of the channels and to modify the roles of the source/sink characteristics of the nodes. We consider the following updating algorithms:
\begin{enumerate}
    \item \texttt{Random pair}: At each iteration, a source (with $S_i = I_0$) and a sink (with $S_j = -I_0$) are randomly chosen from among the $n$ sites.
    \item \texttt{Random half}: At each iteration, $n/2$ sources (each with $S_i = I_0/(\frac{n}{2})$) and $n/2$ sinks (each with $S_j = -I_0/(\frac{n}{2})$) are randomly selected from the $n$ sites; if $n$ is an odd number, one site is considered inactive ($S_j=0$).
    \item \texttt{Random random}: At each iteration, $1 \leq n_{so} \leq n - 1$ sources are randomly selected from the $n$ sites (each with $S_i = I_0/n_{so}$), and the remaining $n-n_{so}$ sites act as sinks (each with $S_j = -I_0/(n-n_{so})$).
    \item \texttt{Random source}: At each iteration, a source (with $S_i = I_0$) is randomly chosen from the $n$ sites, while the remaining $n - 1$ sites function as sinks (each with $S_j = - I_0 / (n-1)$).
\end{enumerate}

Diverse initial conductivity values result in unique steady-state graph configurations or tree structures. A stochastic search yields an array of trees differing in shape and length. All simulations were performed $N_{\text{runs}}$ times, each with distinct random initial conditions for each configuration containing $n$ sites.

The adaptive H-P flow equations require at least one source node and one sink node for the formation of a tree structure. In this study, in a configuration comprising $n$ sites, each site is considered equivalent to every other, rendering each site equally probable to serve as either a source or a sink. Various stochastic algorithms may be utilised to determine the nodes functioning as sources and sinks at each iterative step.

\section{Optimisation criteria}\label{sec4}

Two metrics will characterise each steady-state subgraph: its length $L$, and the average length $\nu$ between all pairs of sites, $(i,j)$.

The length $L$ of a graph is defined by  
\begin{equation} \label{eq:L}
    L = \sum_{(i,j)\in E'} L_{ij},
\end{equation}
where $E'$ denotes the set of edges that are effectively conductive; specifically, these are edges with conductivity values exceeding a predefined threshold, expressed as $E' = \{(i,j)\in E: D_{ij} > D_{\text{thresh}}\}$, where $D_{\text{thresh}} = 10^{-4}$. 

The average length $\nu$ between pairs of sites is specified by 
\begin{equation} \label{eq:nu}
    \nu = \frac{1}{\left(^n_2\right)} \sum^{\left(^n_2\right)}_{(i,j)\in E'} L_{s_i s_j} = \frac{2}{n(n-1)}\sum^{\left(^n_2\right)}_{(i,j)\in E'} L_{s_i s_j},
\end{equation}
where $L_{s_i s_j}$ denotes the shortest path length between sites $i$ and $j$, given that $D_{s_i s_j} > D_{\text{thresh}}$.

From the perspective of the user, an ideal subgraph that interconnects all sources and sinks should demonstrate a low value of $\nu$. In contrast, a subgraph with a low $L$ entails reduced construction and maintenance expenses.

Our objective is to develop algorithms that assist in making decisions regarding the design of communication networks, including roads, railways, and fibre networks.

\section{Results}

To examine the algorithms and parameters described, the simplest case considered is a network represented by a graph embedded in two-dimensional Euclidean space, resulting from a Delaunay triangulation. The nodes form a square point lattice, with positions randomly perturbed by Gaussian noise with a standard deviation of $\sigma = 0.5$. The square point lattice has a side length of $1$ and contains $N = N'\times N'$ nodes. Among these nodes, we select $n$ nodes to serve as sources and sinks.

The parameters for the adaptive incompressible H-P flow are: $\beta = \alpha = 1$, $I_0 = 1$, $V = 100$, $N' = 25$, $\Delta t = 0.1$, $N_{\text{runs}} = 300$.

The cases with  $n=\{3, 4, 5\}$ sources and sinks will be examined, with the sources and sinks arranged in a point lattice within the unit square. Furthermore, a comprehensive analysis of the Portuguese railway system's transportation network will be conducted using a point lattice region shaped like the country's map.

\begin{figure}
 \centering
  \includegraphics[width=0.5\textwidth]{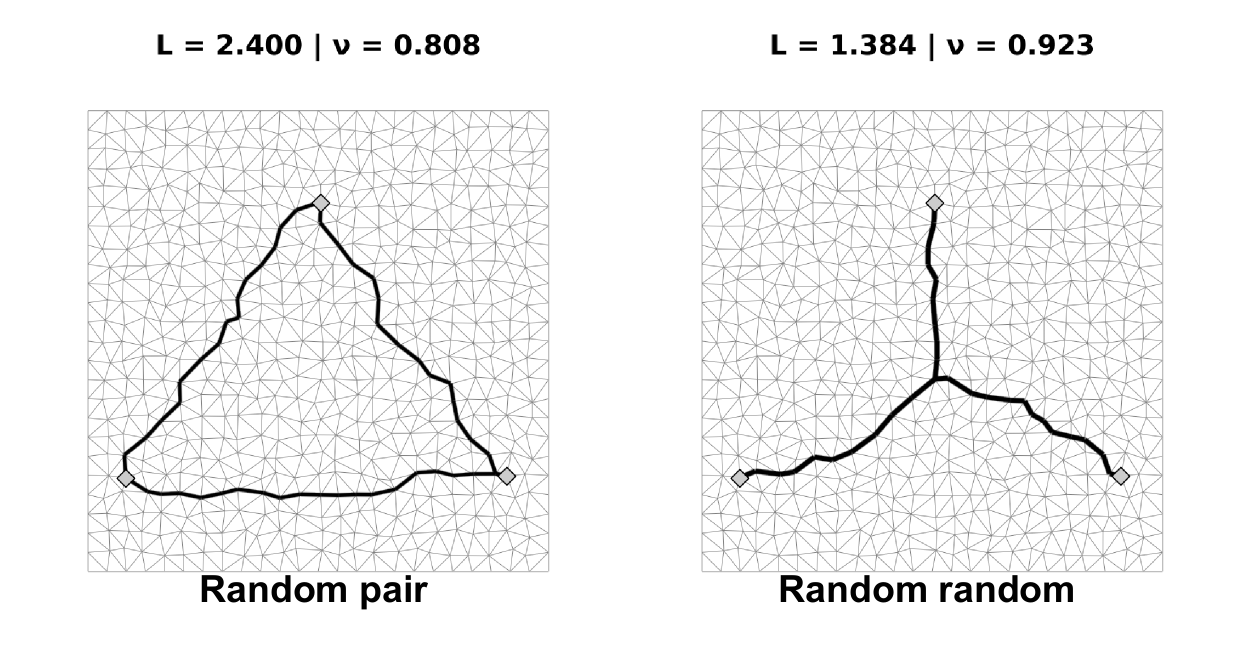} 
\caption{Steady-state  optimised subgraph of the H-P flows for the triangular configuration obtained using the \textbf{Random pair} and \textbf{Random random} algorithms. The optimal communication network is characterised by the minimal value of $\nu$. Additionally, we present the initial graphs $\mathcal{G}$ derived from the Delaunay triangulation of the point lattice within the unit square. The widths of the channels in the steady-state subgraphs are proportional to the adapted channel radius. As the total number of edges in $\mathcal{G}$ approaches infinity, the edges connecting the sources and sinks tend to converge to straight lines.}
\label{fig:tri_random}
\end{figure}

\subsection{Triangular configuration}

For $n = 3$, we selected sites arranged at the vertices of an equilateral triangle with side lengths of $\ell=0.8$. The perimeter of the triangle is $P_\triangle = 2.4$, and the length of the minimal Steiner tree is 
$L_{\text{Stei}_\triangle} = 1.386$.

The steady-state results for the \texttt{Random pair} and \texttt{Random random} algorithms are presented in figure~\ref{fig:tri_random}. The optimal configuration for a communication system is the  \texttt{Random pair} solution. The minimal Steiner tree configuration was obtained using the \texttt{Random random} algorithm.

\subsection{Square configuration}

For $n = 4$, we chose locations configured in a square with a side length of $\ell=0.7$. The perimeter of the square is $P_{\square} = 2.8$, and the minimum Steiner tree length is $L_{\text{Stei}_\square} = 1.912$.

\begin{figure}
\centering
 \includegraphics[width=0.75\textwidth]{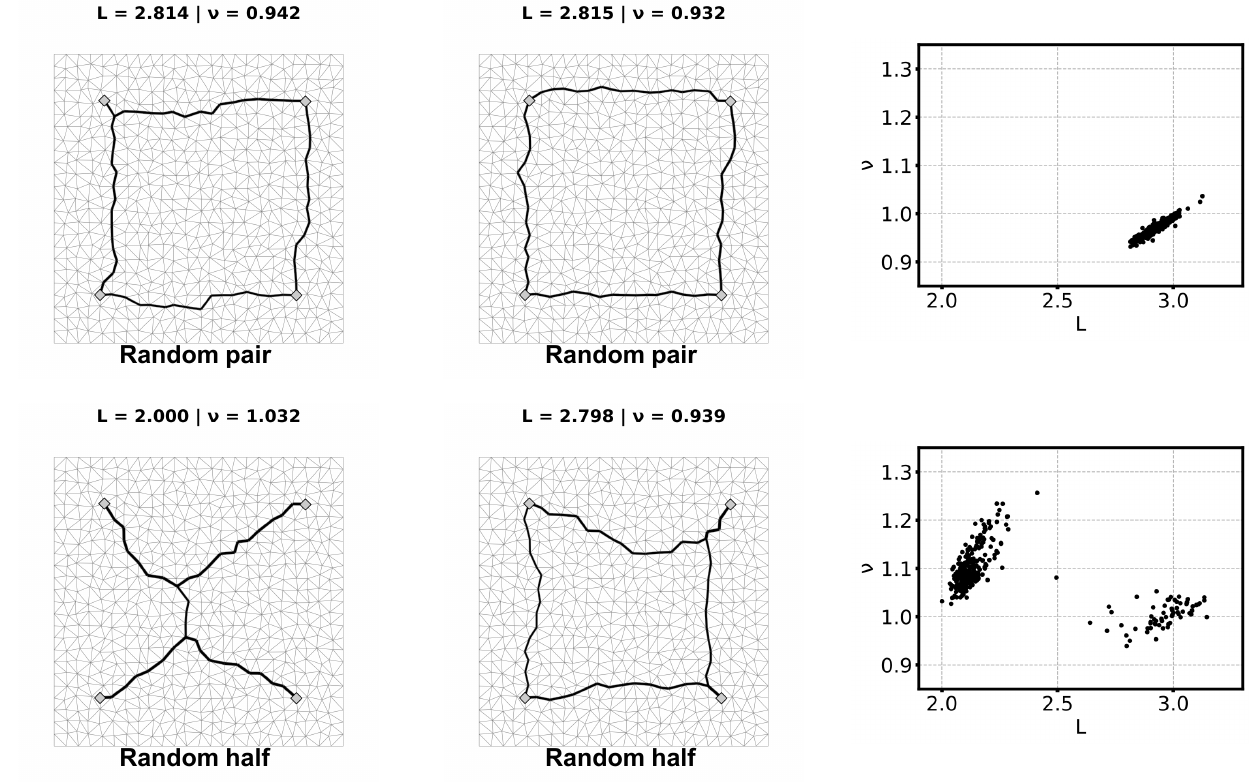}
 \caption{Steady-state optimised subgraph of the H-P flows for the quadrangular configuration obtained using the \textbf{Random pair} and \textbf{Random half} algorithms. The graphs on the right-hand side illustrate the distributions of the parameters $\nu$ and $L$ for the $N_{\text{runs}} = 300$ steady-state solutions obtained with both algorithm.}
\label{fig:square_random}
\end{figure}

The results for the \texttt{Random pair} and \texttt{Random half} algorithms are presented in figure~\ref{fig:square_random}. The displayed outcomes depict the steady states attained with the minimal values of $L$ and $\nu$ for each respective algorithm. We also show the distribution $\nu$ and $L$ for all $N_{\text{runs}} = 300$ runs conducted for each algorithm.

The \texttt{Random pair} algorithm did not provide a solution topologically equivalent to a Steiner tree. The \texttt{Random half} algorithm reveals two suboptimal Steiner-like solutions. Nevertheless, the geometry with lower communication costs from the user's perspective is the square.

\subsection{Pentagonal configuration}

For $n = 5$, we selected sites arranged in a pentagonal formation with sides of length $\ell=0.55$. The perimeter of this pentagon is $P_{\pentago} = 2.75$, and the minimum Steiner tree length is $L_{\text{Stei}_{\pentago}} = 2.14$.

\begin{figure}
\centering
 \includegraphics[width=0.75\textwidth]{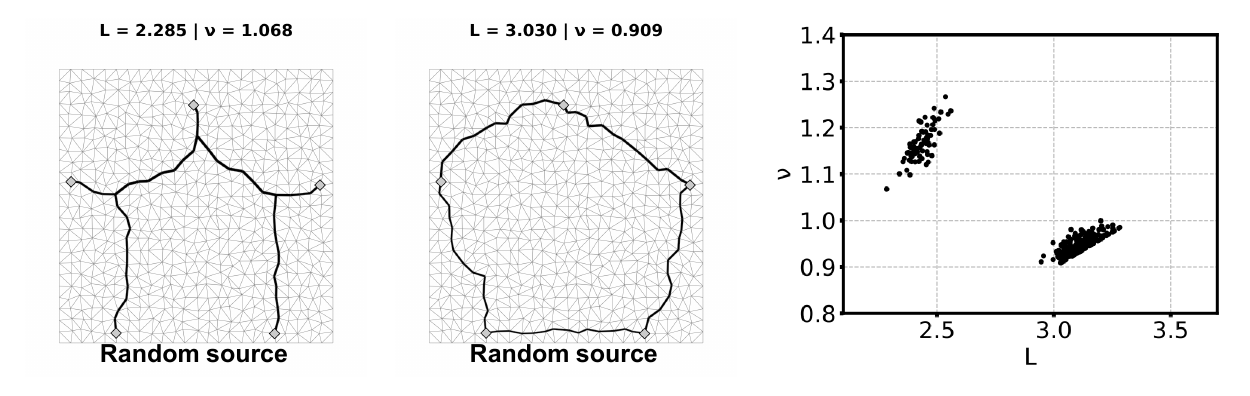}
 \caption{Steady-state optimised subgraph of the H-P flows for the pentagonal configuration obtained with the \textbf{Random source} algorithm. The graph on the right-hand side shows the distributions of the parameters $\nu$ and $L$ for the $N_{\text{runs}} = 300$ steady-state solutions obtained with both algorithm.}
\label{fig:penta_random source}
\end{figure}

The outcomes for the \texttt{Random source} algorithm are illustrated in figure~\ref{fig:penta_random source}. These results depict the steady states attained with the minimal values of $L$ and $\nu$. The distributions of the parameters $\nu$ and $L$$\nu$ for the $N_{\text{runs}} = 300$ steady-state solutions obtained with each algorithm are also shown.

The tree with the smallest $\nu$ for these three algorithms has a length that approximates the perimeter of the square. The \texttt{Random source} algorithm also generated a Steiner tree-like solution, but with a larger value of $\nu$.

\subsection{Portugal communication systems}

The algorithms described will be used to approximate the railway transport network of mainland Portugal. The mesh closely resembles the geographic layout of mainland Portugal and consists of 1005 nodes and 2817 edges, derived through Delaunay triangulation. We identified twenty-five sites of importance, including the eighteen district capitals and seven additional cities of relevance. The length $L_{ij}$ of each edge was determined by the geodesic distance between nodes $i$ and $j$, calculated approximately via the Haversine formula \cite{rodrigo_tese}.

\begin{figure}
\centering
\includegraphics[width=0.88\textwidth]{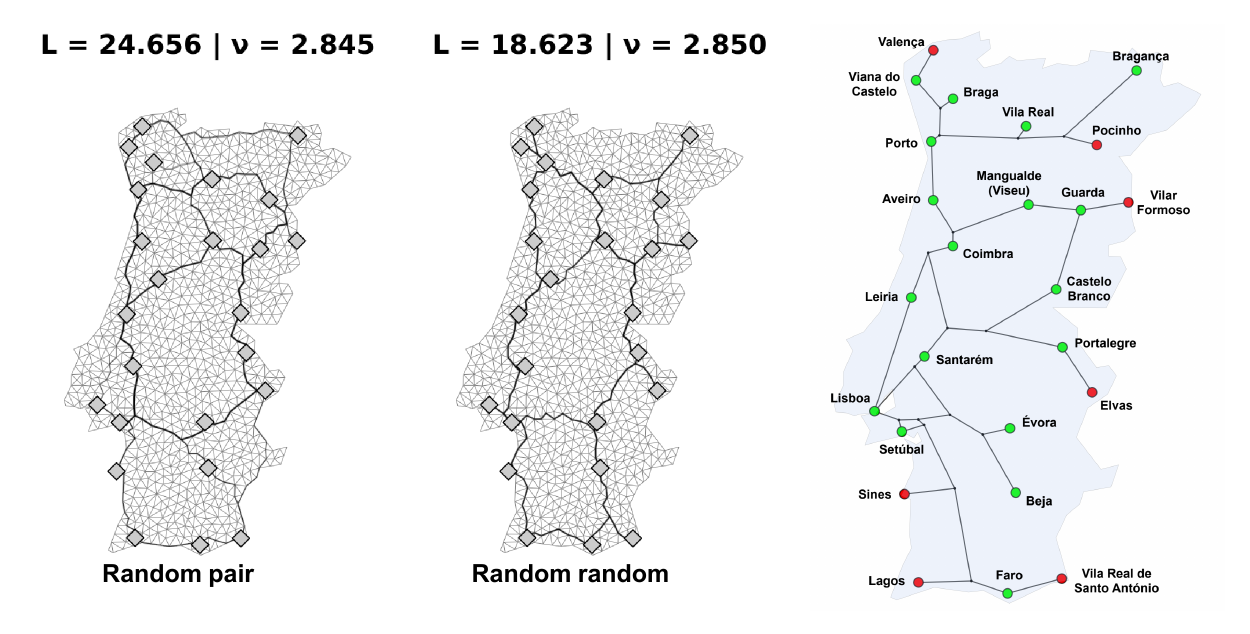} 
\caption{Steady-state the optimised subgraph of the H-P flows for the Portuguese railway system, obtained with the \textbf{Random pair} and \textbf{Random random} algorithms, is to be compared with the existing Portuguese railway network. For the existing network, $L=18.49$ and $\nu=3.154$. For example, comparing the \textbf{Random random} solution with the existing network shows that the H-P steady solution has lower transportation costs and provides greater redundancy at the edge local damage.}
\label{fig:Portugal_rail}
\end{figure}

The results for the \texttt{Random pair} and \texttt{Random random} algorithms are presented in figure~\ref{fig:Portugal_rail} for the steady states obtained with the smallest values of $L$ and $\nu$ values. The smallest $\nu$ tree was generated utilising the \texttt{Random pair} algorithm, with the \texttt{Random random} algorithm closely following. The geometry predicted by the \texttt{Random pair} algorithm demonstrates greater efficiency compared to the existing railway network. The geometries produced by the adaptive incompressible H-P flow algorithm could be further refined, especially concerning the inclusion of Steiner-type points.

\section{Conclusions}

We utilised a formalism describing adaptive incompressible Hagen-Poiseuille flows within channel networks in $\mathcal{N}$-dimensional Euclidean spaces to identify graphs that connect a finite set of nodes. These graphs can represent communication systems and Steiner tree problems. The networks are embedded within a predefined graph $\mathcal{G}$ obtained through Delaunay triangulation from an $\mathcal{N}$-dimensional point lattice. The area of the Delaunay triangles can be sufficiently small to facilitate the system's approach to a continuous limit. This embedding renders the adaptation process uni-dimensional along the edges of the graph. The method is computationally efficient, as multiple replicas for various initial conditions can be generated, enabling the selection of suboptimal or optimal subgraphs and trees, which algebraic techniques can further optimise. This methodology is particularly effective as a practical approach for identifying approximate solutions to Steiner tree configurations.

We have implemented this adaptive incompressible Hagen-Poiseuille formalism in the design of communication systems or vascular networks within organisms. The resulting solutions are either suboptimal or optimal, aimed at minimising the lengths of node connections or transportation costs. These solutions are derived from a sufficiently extensive set of steady-state solutions of the adaptive incompressible Hagen-Poiseuille flows. Additionally, the suboptimality of these solutions underscores the adaptability and resilience of graph networks in response to perturbations. The methodology for designing communication systems presented herein provides a complementary and computationally efficient algorithm for exploring and developing adaptive network routing \cite{kel}.

\end{document}